\documentclass{article}

\usepackage{amsmath, amssymb}

\usepackage[style=numeric, sorting=none, doi=true, maxcitenames=3, mincitenames=1]{biblatex}
\usepackage{authblk}
\usepackage[a4paper, total={6in, 9in}]{geometry}
\usepackage{hyperref}
\usepackage{url}
\hypersetup{colorlinks=true,citecolor=blue,urlcolor=blue,linkcolor=blue}

\usepackage{algorithm}
\usepackage{algpseudocode}

\usepackage{tikz}
\usepackage{float}
\usepackage{graphicx}
\usepackage{caption}
\usepackage{subcaption}
\usepackage{xcolor}
\usepackage{multicol}

\renewenvironment{abstract}{%
	\begin{center}
		\begin{minipage}{0.75\textwidth}
			\footnotesize
			\textbf{Abstract: }
		}{%
		\end{minipage}
	\end{center}
	\vspace{5mm}
}

\title{Optimised Graph Convolution for Calorimetry Event Classification}

\author[1]{Matthieu Melennec\footnote{mmelennec@llr.in2p3.fr}}
\author[1]{Shamik Ghosh}
\author[1]{Fr\'ed\'eric Magniette}

\affil[1]{Ecole Polytechnique, IN2P3-CNRS, Laboratoire Leprince-Ringuet, F-91120 Palaiseau, France}

\date{}

\begin{document}
	\maketitle
	
	\begin{abstract}
		In the recent years, high energy physics discoveries have been driven by the increasing of luminosity and/or detector granularity. This evolution gives access to bigger statistics and data samples, but can make it hard to process the detector outputs with current methods and algorithms. Graph convolution networks, have been shown to be powerful tools to address these challenges. We present our graph convolution framework for particle identification and energy regression in high granularity calorimeters. In particular, we introduce our algorithm for optimised graph construction in resource constrained environments. We also introduce our implementation of graph convolution and pooling layers. We observe satisfying accuracies, and discuss possible application to other high granularity particle detector challenges.
	\end{abstract}

    \section{Introduction}
    
    	The increase in luminosity at HL-LHC \cite{zurbano_20} and detector granularity for the CMS HGCAL upgrade \cite{cms_17} will allow to observe complex phenomena at an increased level of precision. This will increase the available statistics and precision by multiple orders of magnitude, which facilitates the detection of rare events, at the price of a significant increase of the number of channels. The challenge is that most of the standard techniques for reconstruction and triggering are not operative in such a context. For example, the energy threshold-based triggers fail to handle the complexity of high pile-up collisions. While deep learning methods such as convolution techniques are known to handle noisy and complex data inputs well \cite{li_21}, they do not generalise well to the very peculiar topologies of particle detectors. Alternatives such as spatial graph convolution have been shown to handle well such data \cite{zhang_22}. In particular, they have proven to give excellent results on particle detector data at LHC \cite{qasim_19} and neutrinos experiments \cite{choma_18} for tasks such as classification, energy regression and object segmentation.
    	We present the use of graph convolution neural networks for calorimeter object identification and energy regression, in the context of the OGCID project \footnote{Optimisation of Graph Convolution for particle IDentification: \url{https://llrogcid.in2p3.fr/}}. In particular, we use the knowledge of the detector geometry to pre-compute proximities of potential node positions in the data to optimise the graph construction procedure, and discuss its time complexity. We also present the chosen message passing convolution operation, as well as introduce our pooling algorithm, Treclus, to coarsen local substructures, while preserving the overall structure of the graph. We present the graph readout pooling operation, used to flatten the outputs of convolution to an MLP of constant size. Finally, we present the classification and energy regression capacities of our pipeline, looking in particular at the resolution in energy to check the validity of our approach.
    	
    \section{The D2 Dataset}
    
    	The data used in this study are extracted from the OGCID/D2 simulation dataset \cite{becheva_24}. It is a set of single particle interaction with a simplified high granularity calorimeter inspired by CMS HGCAL \cite{cms_17}.
    	This detector architecture features two main sections: the electromagnetic calorimeter (ECAL) and the hadronic calorimeter (HCAL), optimised for detecting particles of varying energies. The ECAL comprises 26 layers of lead absorbers (6.05	mm thick), while the HCAL has 24 layers of stainless steel absorbers, with 12 layers of 45 mm thickness and 12 of 80 mm, providing high resolution sampling of particle showers. Active silicon layers are placed between absorbers, segmented into hexagonal cells with a thickness of 0.32 mm and a transverse area of approximately 1 cm$^2$ for ECAL and 4 cm$^2$ for HCAL. The ECAL layers are comprised of 3571 Si sensor cells for a total width of 680 mm, and the HCAL layers have 1801 Si cells and are 960 mm wide, for a total of 136,070 channels\footnote{\url{https://llrogcid.in2p3.fr/hgcal-like-simulations/}} (versus 6 million for HGCAL \cite{cms_17}). 
    	The D2 dataset is a collection of simulated interactions between this detector and four types of particles (muon, positively charged pions, electrons and photons) at different energies and with a flat incidence angle (aligned with the longitudinal detector axis). The dataset includes detailed informations on each particle event, namely the energy measurement in each cell but also extracted variables of interest describing the geometry and the energy repartition of the hits.
    	For this study, we looked at electrons, photons, muons and pions with energies ranging from 10 to 100 GeV and aligned with the longitudinal axis. The event data that was used was the energy measurement in each cell.
    	\begin{figure}
    		\centering
    		\begin{subfigure}{0.3\textwidth}
    			\centering
    			\includegraphics[width=\linewidth]{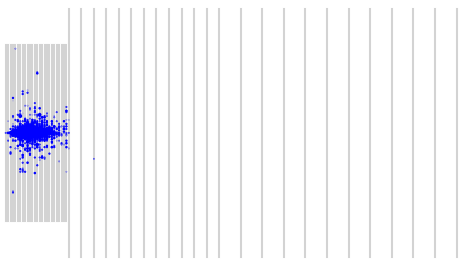}
    			%\caption{}\label{SubFig-event_displays-elm}
    		\end{subfigure}
    		\hfill
    		\begin{subfigure}{0.3\textwidth}
    			\centering
    			\includegraphics[width=\linewidth]{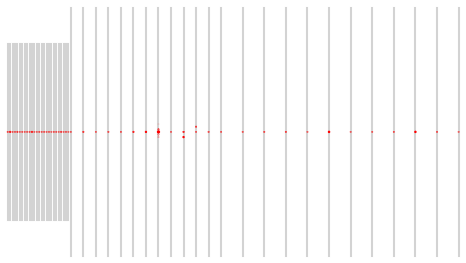}
    			%\caption{}\label{SubFig-event_displays-mum}
    		\end{subfigure}
    		\hfill
    		\begin{subfigure}{0.3\textwidth}
    			\centering
    			\includegraphics[width=\linewidth]{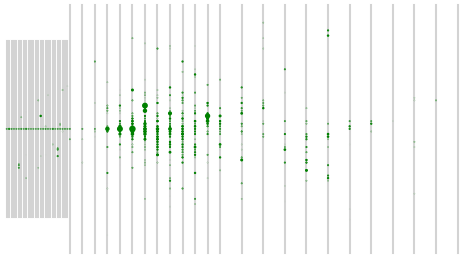}
    			%\caption{}\label{SubFig-event_displays-pip}
    		\end{subfigure}
    		\vspace{-0.4cm}
    		\caption{Event displays for an electron/photon shower (blue), a muon (red), a pion shower (blue). Photon showers are very similar to electron showers \cite{wigmans_17}.}\label{Fig-event_displays}
    	\end{figure}
    	
    \section{Code Architecture}
    
    	The pipeline used in this work is based on message passing convolution \cite{gilmer_17}. After generating graphs from the raw event data, these go through several layers of message passing convolution and pooling (CP layers). The resulting graphs are then flattened into a one-dimensional tensor by a readout pooling layer, which flattens the node features of the graphs in an orderly fashion. It also returns a structured tensor of constant size, independent of the input graph’s size. This layer is essential to then feed the convoluted features to a multi-layer perception that will output our objective (particle ID or energy) by following a systematic pattern.
    	
    	\subsection{Graph Construction and Proximity Tables}
\begin{figure}
    			\centering
    			\begin{minipage}{0.45\textwidth}
					\centering
					\begin{tabular}{c||cccc}
						Sensor id & pos. 1   & pos.2    & $\cdots$ & $k$      \\ \hline \hline
						1         & $s_1^1$  & $s_1^2$  & $\cdots$ & $s_1^k$  \\
						%2         & $s_2^1$  & $s_2^2$  & $\cdots$ & $s_2^k$  \\
						$\vdots$  & $\vdots$ & $\vdots$ & $\ddots$ & $\vdots$ \\
						i         & $s_i^1$  & $s_i^2$  & $\cdots$ & $s_i^k$
					\end{tabular}
					\vspace{-0.3cm}
					\captionof{table}{Proximity table. $s_i^k$ is the sensor in $k$-th position in the row for sensor $s_i$. Neighbours are ordered such that for all $i,k, d(s_i, s_i^k) \leq d(s_i, s_i^{k+1})$.}
					\label{Table-PT}
					\includegraphics[width=\textwidth]{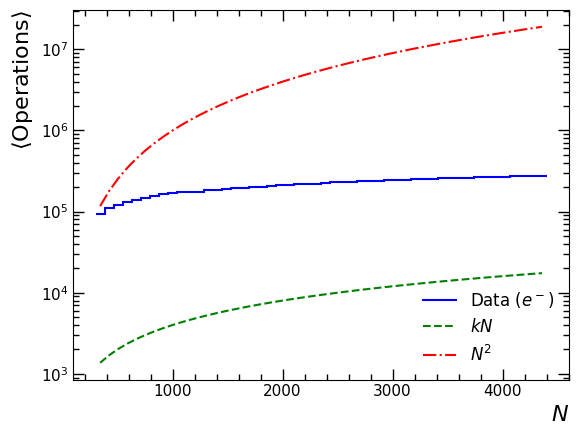}
					\vspace{-0.3cm}
					\captionof{figure}{Average number of operations for the PT-KNN algorithm. A comparison shows linear and quadratic complexities. Asymptotically, the average number of operations behaves like $\mathcal{O}(kN)$}
				\end{minipage}
				\hfill
				\begin{minipage}{0.45\textwidth}
					\captionof{algorithm}{PTKNN: Proximity table $k$ nearest neighbours graph generation algorithm. $k$ is the number of neighbours wanted for each node.}
					\label{Algo-PTKNN}
					\begin{algorithmic}[1]
						\State \textbf{Input}: hits, $k$
						\State \textbf{Output}: $G=\{V, E\}$
						\State $V \leftarrow \{\}$
						\State $E \leftarrow \{\}$
						\For{$s_i \in$ hits}
							\State $V \leftarrow V \cup \{s^i\}$
							\State counter $\leftarrow$ 0
							\While{counter $< k$}
								\For{$s_i^k \in \{s_i^k\}_k$}
									\If{$s_i^k \in$ hits}
										\State $V \leftarrow V \cup \{s_i^k\}$
										\State $E \leftarrow E \cup \{(s_i, s_i^k)\}$
										\State counter $\leftarrow$ counter $+ 1$
									\EndIf
								\EndFor
							\EndWhile
						\EndFor
						\State \textbf{return} $G = \{V, E\}$
					\end{algorithmic}
				\end{minipage}
			\end{figure}

\begin{figure}
    			\centering
    			\begin{subfigure}{0.48\textwidth}
    				\centering
    				\includegraphics[width=0.8\textwidth]{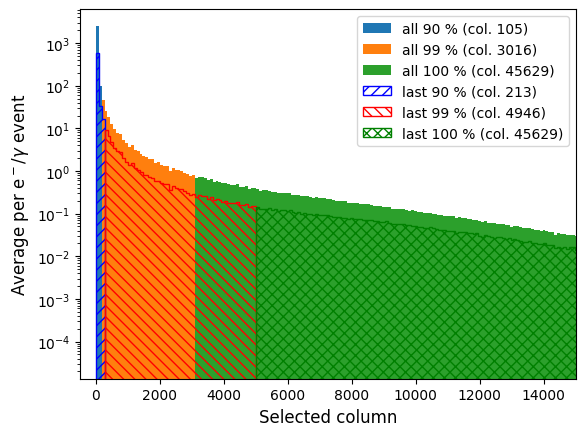}
    				%\caption{Selected columns for e$^-$/$\gamma$ events}
    			\end{subfigure}
    			\begin{subfigure}{0.48\textwidth}
    				\centering
    				\includegraphics[width=0.8\textwidth]{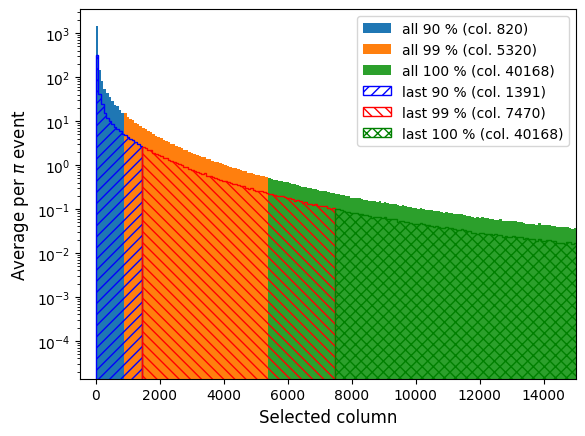}
    				%\caption{Selected columns for $\pi^+$ events}
    			\end{subfigure}
    			\vspace{-0.3cm}
    			\caption{Column selections for PT-KNN algorithm with $k=4$, euclidean distance at energies 10 to 100 GeV.}\label{Fig-PT_cut}
    		\end{figure}

    		Every calorimeter event is represented as a graph, the nodes of which correspond to every hit cell of the event, linked by arbitrarily defined edges. Our edge building strategy uses the $k$ nearest neighbours (KNN) algorithm to encode a sense of geometric locality in the graph structure. Our implementation of KNN is based off of so-called proximity tables (PTs), which contain a pre-computed order of neighbourhood for every sensor of the detector. That is, for every sensor of the detector, all other sensors are ordered according to a user-defined metric, as in Table \ref{Table-PT}. The graph building procedure then follows algorithm \ref{Algo-PTKNN}.
    		Note that while it is standard to use the euclidean distance for the KNN algorithm, other metrics can be used to encode physical properties in the graph structure.
    		%For instance, one could use the correlation of cell hits as a metric, or connect the outermost nodes with the centre of the shower by introducing a radiality term. 
    		In this study, the choice of the metric  seemed not to have an impact on the performance of the reconstruction, and we therefore only use the euclidean distance. We also use the full proximity table for the D2 detector, i.e. $m^2$ entries with $m$ the number of Si sensors in the detector.
    		Intuitively, one understands that the PT-KNN algorithm brings an improvement to the KNN algorithm with regards to time complexity. Indeed, the KNN complexity, as used here for graph construction where for every nodes, we look through all the other nodes for the $k$ nearest, is constantly quadratic. For the PT-KNN algorithm, while the worse case complexity can in theory increase to $\mathcal{O}(NM)$, with $N$ the number of nodes and $M$ the number of cells in the detector if $M \gg N$, this risk is mostly mitigated by the geometric locality of the shower (figure \ref{Fig-event_displays}). In practice, in the range of number of nodes $N$ corresponding to the size of a shower in the detector, the observed average last selected column $\left\langle c_{\text{last}} \right\rangle$ gives an average complexity $N \left\langle c_{\text{last}} \right\rangle \sim (\log N)^2$. To verify this, we assume the nodes are spread following a normal distribution and combine it with the number $N_\text{cells} \propto r$ of cells found in a radius $r$ around the Gaussian's mean. This allows us to recover the radius where on average $k$ nodes can be found and estimate $\langle c_\text{last} \rangle = A\log^2 N \log\left(N/(N-k)\right) + k$, which when fitted to the data gives $(\log N)^2/N$ at leading order\footnote{A fully detailed proof may be found on the \href{https://llrogcid.in2p3.fr/graph-construction-complexity/}{OGCID project web page}}. In the asymptotic range $N\to\infty$ where $N \gg \left( \log N \right)^2$, one must consider the lower bound $kN$ imposed by the PT-KNN algorithm, and we obtain a linear asymptotic complexity.
    		The value of $\left\langle c_\text{last} \right\rangle$ is particularly relevant in the context of constrained computing, such as particle detector triggering algorithms. Indeed, reducing the size of the proximity table would greatly reduce the memory footprint of the PT. Hence knowing how far the PT is explored allows us to truncate the rows to $\lambda \left\langle c_\text{last} \right\rangle$ columns for some $\lambda$ to be determined. In practice, we observe that for most hits, the PT-KNN algorithm only explores 5\% of the corresponding row in the PT (figure \ref{Fig-PT_cut}). Hence a strong truncation of the depth of the PT would likely not have too big an impact on the graph structure while drastically reducing the size of the PT. The truncation of the proximity table also ensures that the worse case complexity reduces to $\mathcal{O} \left( N m_{\text{max}} \right)$ with $m_{\text{max}}$ the number of columns kept with typically $m_{\text{max}} \sim N$.
    		The resulting graphs are composed of vertices $v \in V$, holding as features the energy $x_v$ measured by the corresponding sensor and the position $\vec{u}_v$ of that sensor. Note that in order to respect the symmetries of shower formations in the detector, the position of the node is kept as a ``hidden feature,'' not used during the convolution operation. The positional information of the graph is encoded in the edges $e_{vw} \in E$, which carry as feature $x_{vw} = \left\Vert \vec{u}_v - \vec{u}_w \right\Vert_2$ the euclidean distance between the end nodes $v, w \in V$, holding information about \textit{relative} positions. We observed that when increasing $k$, the improvement in classification and regression power formed a plateau from $k=4$, while being more computationally expensive. As such, All the graphs built in this work use $k=4$.
			    	
    	\subsection{Message Passing Convolution}
    		Message passing graph convolution can be seen as a generalisation of image convolution, were instead of operating on pixels, we operate on the nodes and/or edges of a graph \cite{gilmer_17}. We chose the message function to be a single layer of neurons with a Leaky-ReLU activation function, taking as input $\chi$, the concatenated features $x_v$, $x_w$ and $x_{vw}$:
    		\begin{equation}\label{message}
    			\phi_{\theta_\phi} \left( x_v, x_w, x_{vw} \right) = \text{Leaky-ReLU} \left( \Theta_\phi \chi \right), \quad \chi := \left[ x_v, x_w, x_{vw} \right] \in \mathbb{R}^{2n + m} 
    		\end{equation}
    		where $n$ denotes the number of features per node, $m$ the number of features per edge, and $\Theta_\phi \in \mathbb{R}^{l \times (2n + m)}$ is a matrix of trainable weights, with $l$ the output size of the message function.
    		Messages from all neighbours are aggregated by a feature-wise max pooling for particle IDing problems in order to identify protruding features \cite{goodfellow_16} and mean pooling for energy regression in order to smoothen the protruding features and consider the graph globally \cite{iandola_14}. Finally, we update the features of node $v$ by performing the aggregation of the aggregated messages with the features of the node $v$ itself. In practice, this is equivalent to adding self-loops to the graphs, i.e. for any $v\in V, e_{vv} \in E$.
    		\begin{equation}
    			\gamma_{\theta_\gamma} \left( x_v^{(l)}, \underset{w\in\mathcal{N}(v)}{\square} \phi_{\theta_\phi} \left( x_v^{(l)}, x_w^{(l)}, x_{vw} \right) \right) = \underset{w\in\tilde{\mathcal{N}}(v)}{\square} \phi_{\theta_\phi} \left( x_v, x_w, x_{vw} \right)
    		\end{equation}
    		with $\tilde{\mathcal{N}}(v) = \mathcal{N}(v) \cup \{ v \}$. In our case, the edge features are the euclidean distance between nodes. As such, for any node $v$, $x_{vv} = 0$.
    		
    	\subsection{Graph Pooling}\label{SubSec-pooling}
\begin{figure}
    			\centering
    			\begin{subfigure}{0.3\textwidth}
    				\centering
    				\includegraphics[width=\textwidth]{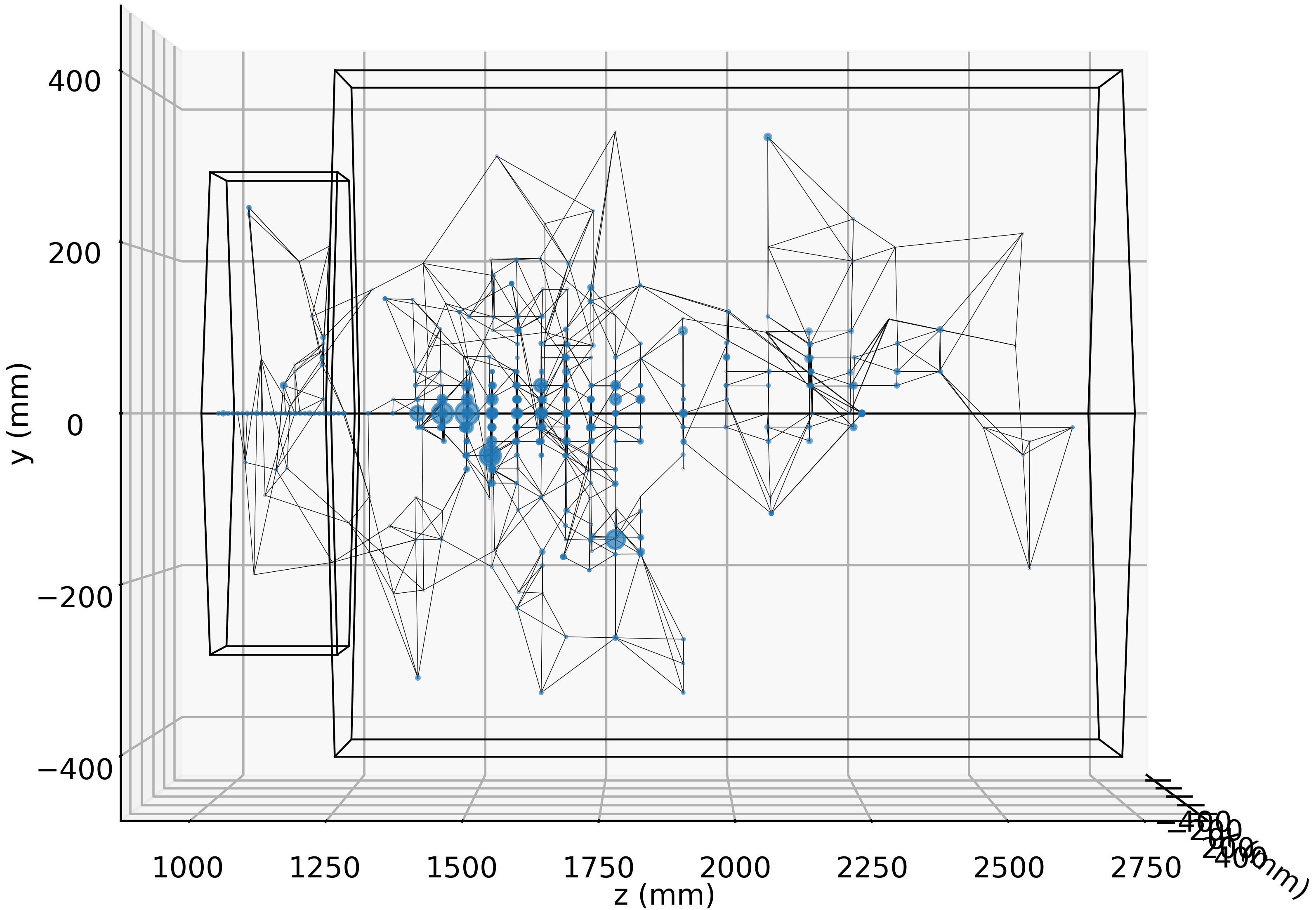}
    			\end{subfigure}
    			\begin{subfigure}{0.3\textwidth}
    				\centering
    				\includegraphics[width=\textwidth]{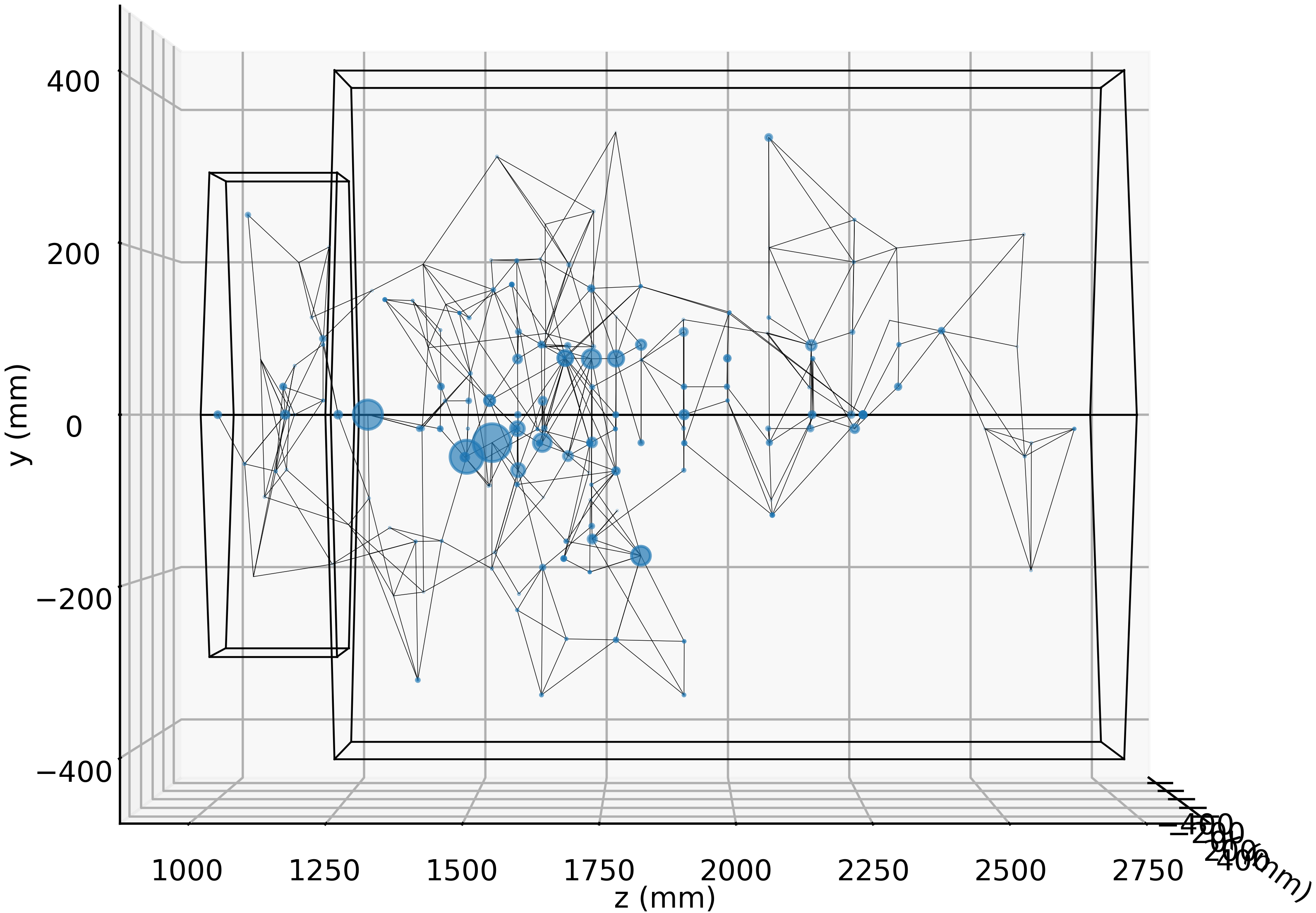}
    			\end{subfigure}
    			\begin{subfigure}{0.3\textwidth}
    				\centering
    				\includegraphics[width=\textwidth]{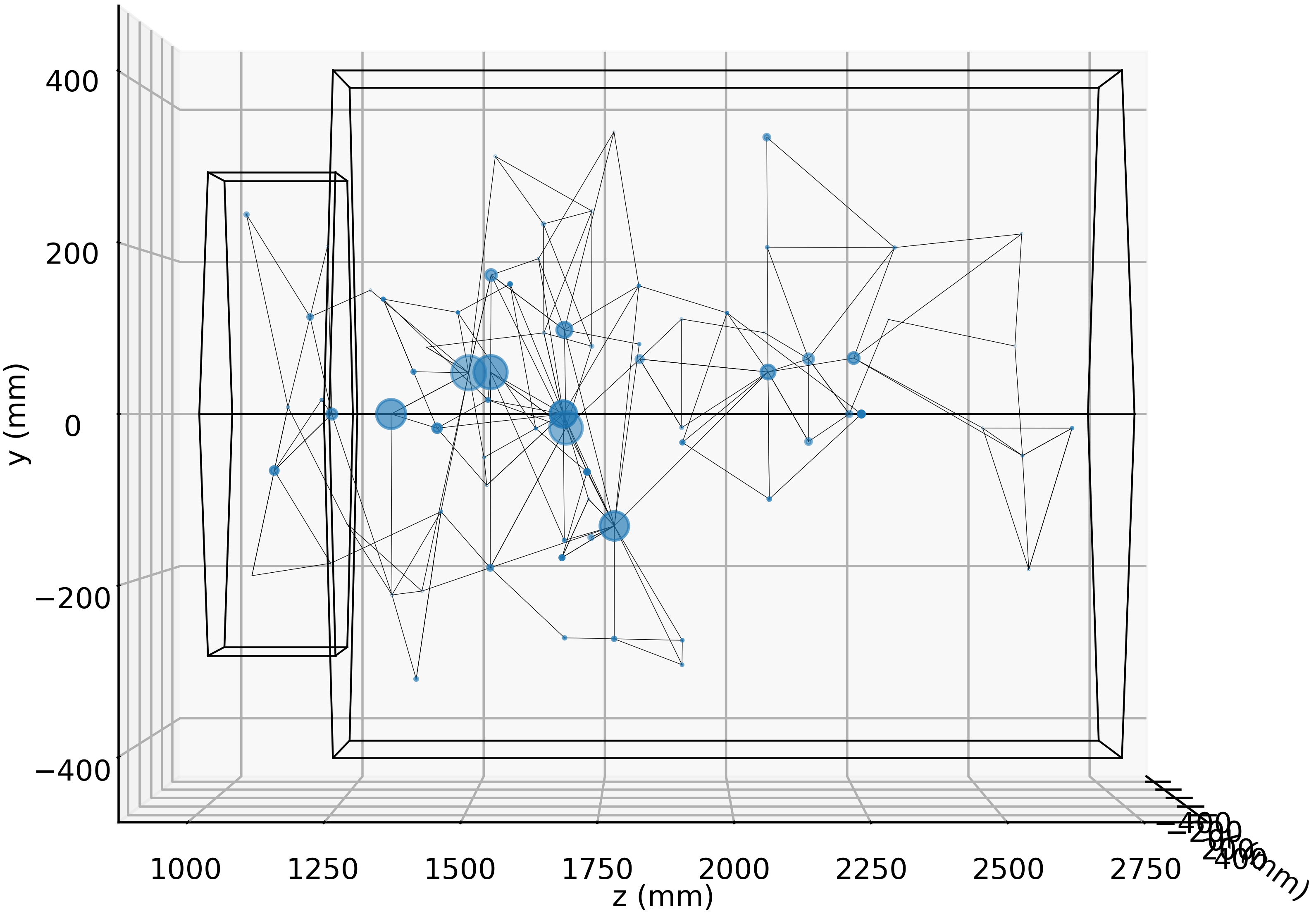}
    			\end{subfigure}
    			\vspace{-0.2cm}
    			\caption{$\pi^+$ shower graph representation after successive pooling steps. We indeed observe that while the number of nodes decreases, the global shape of the graph is preserved through pooling.}\label{Fig-pooling}
    		\end{figure}

    		Graph pooling is an important step in graph convolution, as it allows to coarsen the graph, hence looking at the graph on a more global scale \cite{defferrard_16}. First, let us express the steps of a graph pooling operation as in \cite{grattarola_24}, with a clustering step, where we select which nodes will be clustered together, followed by a reduction step, where nodes within clusters are pooled, and finally a connection step, during which we update the adjacency of clusters. In our case the graphs are embedded in a euclidean space, and we therefore want to cluster together neighbours that are geometrically close to one another, as illustrated in figure \ref{Fig-pooling}.
    		To this end, we developed a threshold clustering algorithm, called Treclus, that creates a matching of the graph \cite{lovasz_09}, where matched nodes are linked by an edge shorter than a threshold $\varepsilon$, and clusters matched nodes together. It is very important that the clustering edges make a matching to keep a linear complexity of the clustering algorithm. This however implies that clusters may only contain two nodes, while we ideally wish to find a maximum matching of the subgraph $G'=(V, E' = \{ e\in E \vert e \leq \varepsilon \})$, which is usually a quadratic problem \cite{behnezhad_24}. Our approach therefore consists in applying the Treclus algorithm multiple times in a row to increase the number of clustered short edges with limited impact on complexity. We observed that less than 10 calls to Treclus suffice to collapse all edges shorter than the 30-th percentile of edges after at most 2 CP layers. 
    		The reduction step consists of a feature wise pooling of nodes within the same clusters. The choice of the pooling operation follows the same logic as the aggregator function choice for the convolution step. However, while this applies strictly for the classification task (with a max pooling), for the energy regression step, the aggregator must be a feature-wise sum to conserve information about the number of nodes contained in the cluster. The position of the cluster is taken at random from one of the nodes it contains. This allows to keep the graph embedded in the known detector geometry (see section \ref{SubSec-readout}) while preserving the locality of the cluster, since the length of the remaining edges makes the induced error negligible.
    		Finally, for the reconnection step, the clusters inherit their adjacencies from their nodes, i.e. two clusters are neighbours if they each contain neighbouring nodes. The length of these edges is then re-computed to smoothen the accumulation of errors due to the random choice of cluster positions.
    		
    	\subsection{Graph Readout Pooling}\label{SubSec-readout}
    		
\begin{figure*}  % Float container
    			\centering
    			\begin{minipage}{0.5\textwidth}
    				\centering
    				\begin{subfigure}{0.45\textwidth}
    					\centering
    					\includegraphics[width=0.6\textwidth]{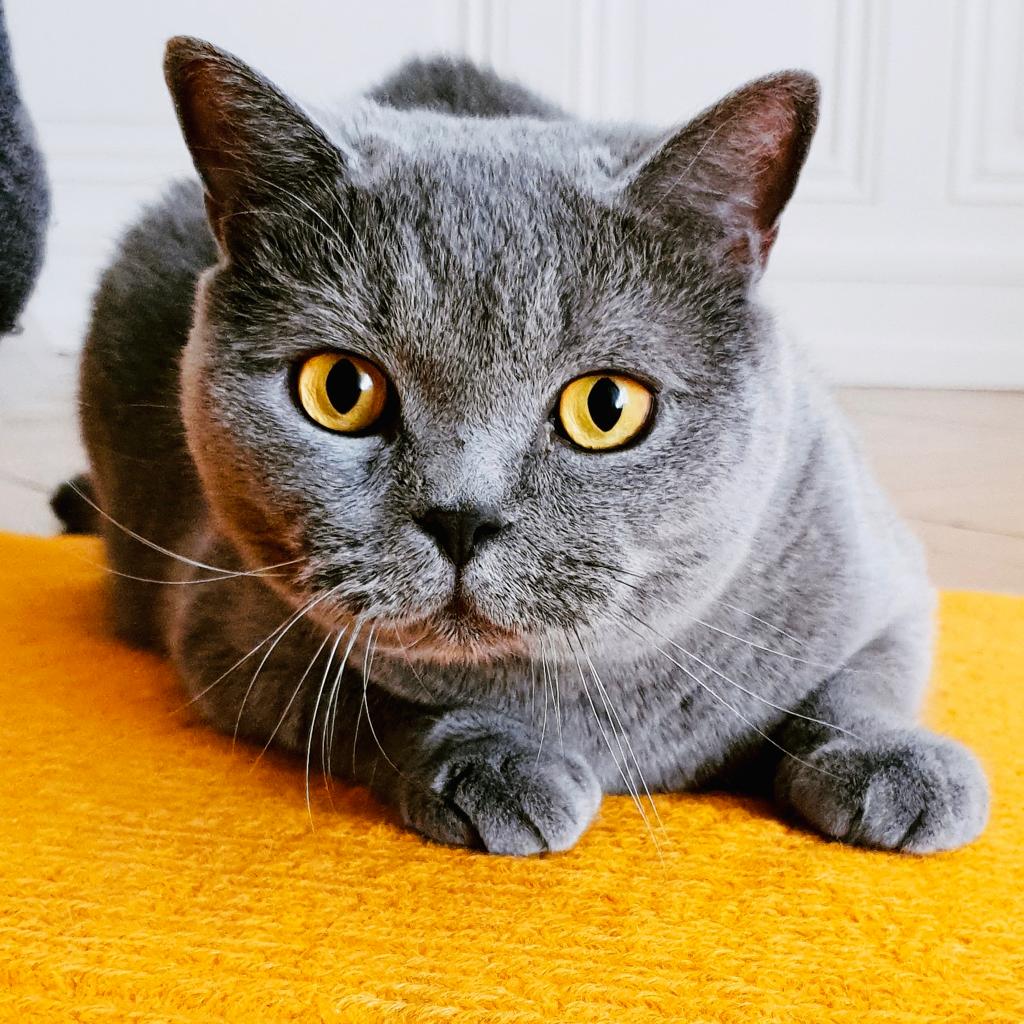}
    					\caption{}\label{SubFig-milo}
    				\end{subfigure}
    				%\hspace{0.5cm}
    				\begin{subfigure}{0.45\textwidth}
    					\centering
    					\includegraphics[width=0.6\textwidth]{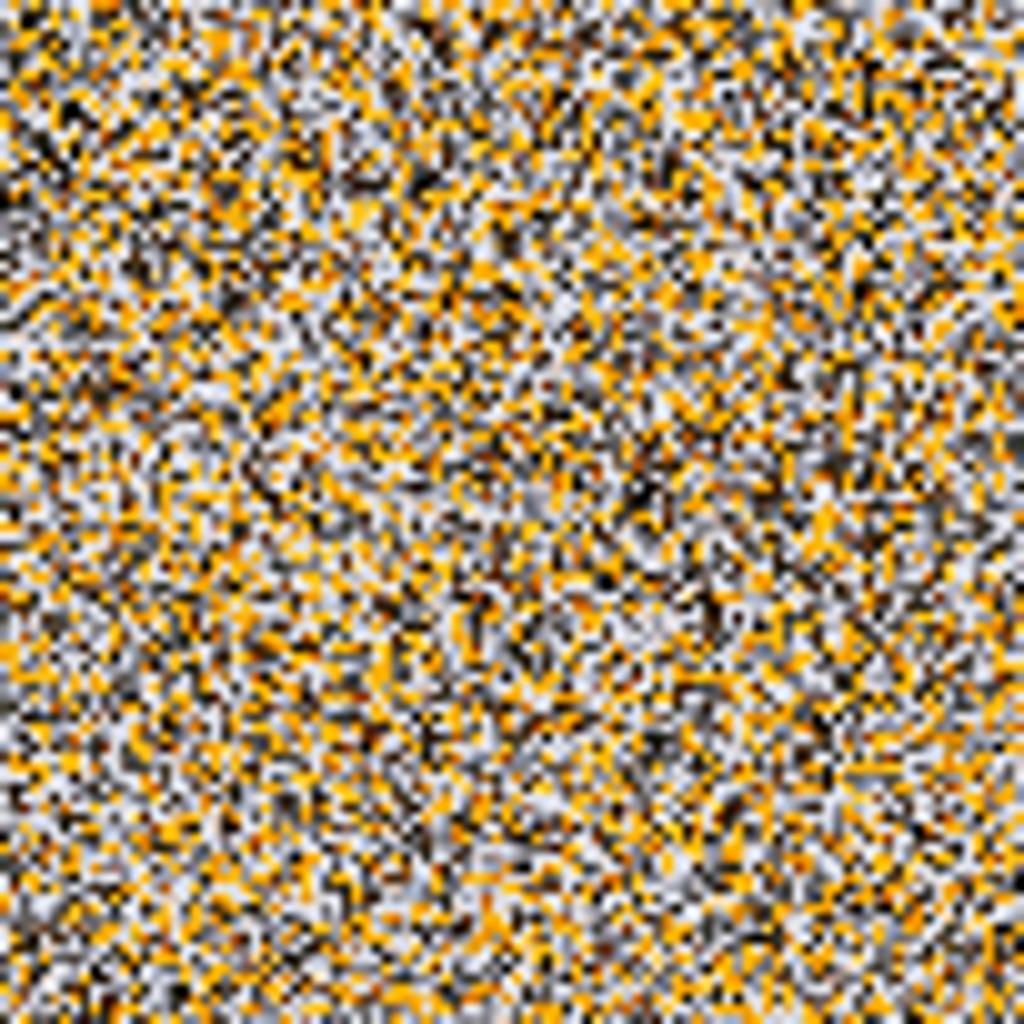}
    					\caption{}\label{SubFig-shuffle}
    				\end{subfigure}
    				\vspace{-0.68cm}
    				\captionof{figure}{While an image with an expected and ordered structure (\ref{SubFig-milo}) is easily interpreted, a random order (\ref{SubFig-shuffle}) makes it uninterpretable.}\label{Fig-milo_shuffle}
    			\end{minipage}
    			\hfill
    			\begin{minipage}{0.45\textwidth}
    				\centering
    				\includegraphics[width=0.45\textwidth]{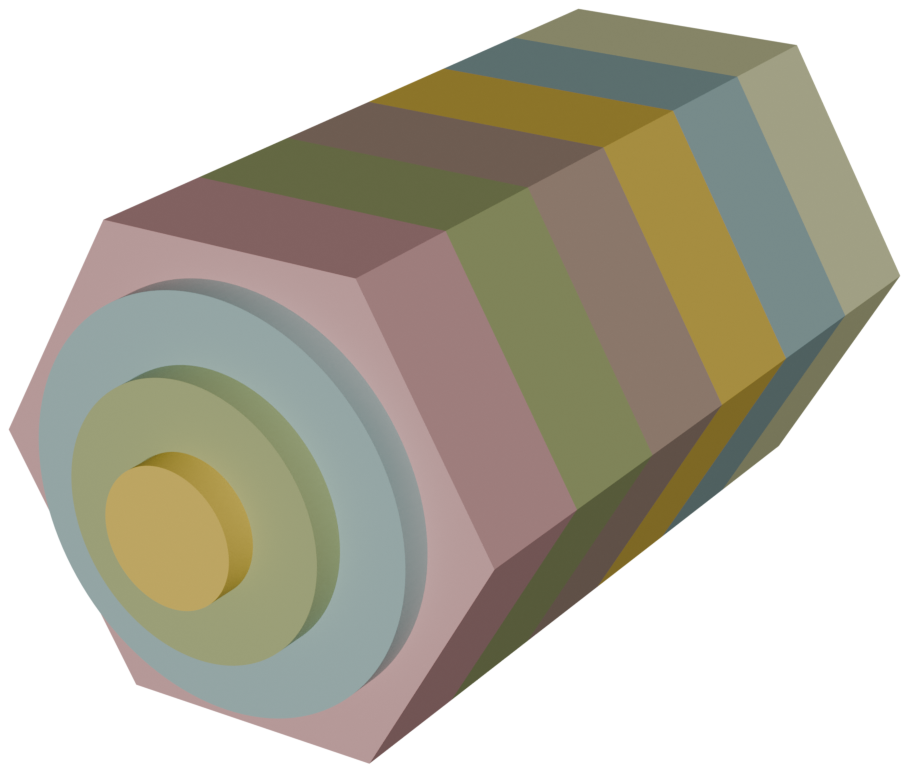}
    				\vspace{-0.25cm}
    				\captionof{figure}{Radial readout regions of the detector respecting the rotational symmetry of the detector.}\label{readout_regions}
	   			\end{minipage}
    		\end{figure*}

    		After the convolution and pooling layers, it is necessary to flatten the graph representation if the output of convolution needs to be fed into a multi-layer perceptron (MLP). This operation, known as readout (or readout pooling), can be quite tricky, as graph-like data usually does not have any consistent ordering and may have variable sizes. The importance of ordering for the readout operation is illustrated in figure \ref{Fig-milo_shuffle}. The same applies to the readout of graphs and the ordering of nodes \cite{zhang_18}.
    		Our approach uses the fact that we know the detector geometry, and kept the nodes embedded (through the hidden features) in that geometry (see ``reduction'' paragraph of Sec. \ref{SubSec-pooling}) to re-embed the graph back into the detector, and divide the detector into different \emph{readout regions}. Using the same argument and procedure as in section \ref{SubSec-pooling}, all nodes within the same readout region are feature-wise pooled together, and the obtained feature of every region are flattened following a known order. It is however important that this voxelisation of the detector respects the geometry of the detector (see figure \ref{readout_regions}). Indeed, the HGCAL model that we considered in this study has a rotational symmetry. If we rotated an electron shower in the detector, while it is the same input, if the readout does not respect that symmetry in the detector the model would have to learn both showers as two different inputs with identical output. This would hinder the training time and performance greatly.    		
    		Note that the detector being composed of two different types of detector, electromagnetic and hadronic calorimeters, we can adapt the granularity of the readout to the task at hand: in energy regression tasks for electrons or photons, where there is no interaction with the hadronic part of the detector, a higher granularity of the readout is only needed in the ECAL.
    	\begin{figure*}
    		\centering
    		\includegraphics[width=0.8\textwidth]{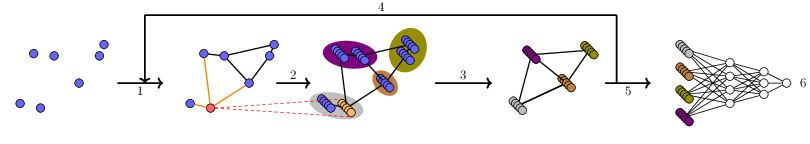}
    		\vspace{-0.2cm}
    		\caption{Typical message passing GNN pipeline. (1) Graph generation (2) Message passing convolution - Nodes collect ``messages'' from their neighbours (3) Pooling (4) Repeat CP layers (5) Readout (6) Objective (PID or Energy regression)}
    	\end{figure*}
    		
    	\subsection{Single Particle Reconstruction Pipeline}
    	
\begin{figure*}
    			\centering
    			\scalebox{0.5}{
    				\begin{tikzpicture}
    					\node[draw, rectangle, anchor=center, inner sep=7pt] (init) at (0,0) {Detector event};
    					\node[draw, rectangle, anchor=center, inner sep=7pt] (graph) at (5,0) {Event graph};
    					\draw[->] (init) -- (graph) node[midway, above] {PT-KNN};
    					
    					\node[draw, rectangle, anchor=center, inner sep=7pt] (PID) at (8.5,0) {Particle ID};
    					\draw[->] (graph) -- (PID);
    					
    					\node[draw, rectangle, anchor=center, inner sep=7pt] (egamma) at (11.5,1) {e$^- / \gamma$};
    					\node[draw, rectangle, anchor=center, inner sep=7pt] (pi) at (11.5,0) {$\pi^+$};
    					\node[draw, rectangle, anchor=center, inner sep=7pt] (mu) at (11.5,-1) {$\mu$};
    					\draw[->] (PID) -- (egamma);
    					\draw[->] (PID) -- (pi);
    					\draw[->] (PID) -- (mu);
    					
    					\node[draw, rectangle, anchor=center, inner sep=7pt] (egammareg) at (15,1) {e$^- / \gamma$ Energy regression};
    					\draw[->] (egamma) -- (egammareg);
    					\node[draw, rectangle, anchor=center, inner sep=7pt] (pireg) at (15,-0.2) {$\pi^+$ Energy regression};
    					\draw[->] (pi) -- (pireg);
    					
    					\node[draw, rectangle, anchor=center, inner sep=7pt] (energy) at (19,0) {Energy};
    					\draw[->] (egammareg) -- (energy);
    					\draw[->] (pireg) -- (energy);
    					
    				\end{tikzpicture}
    			}
    			\vspace{-0.2cm}
    			\caption{Algorithmic workflow of the single particle full reconstruction pipeline}\label{Fig-hibou_pipe}
    		\end{figure*}
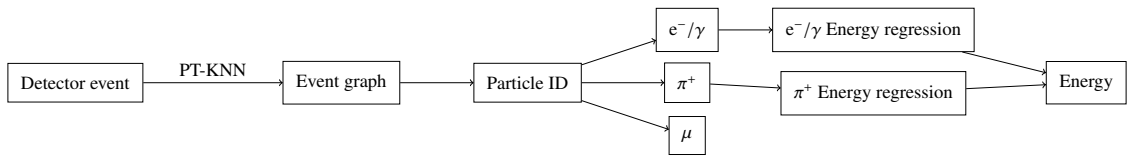

    		All the elements presented above allow us to design a pipeline for the reconstruction (particle ID and energy regression) of calorimeter events. The pipeline aims at identifying the detected particle and regressing its energy. The workflow is illustrated in figure \ref{Fig-hibou_pipe}.
    		The first step is to generate graphs from raw detector data, using the PT-KNN algorithm. A first pipeline is trained for particle classification. Because electromagnetic and hadronic showers have very different signatures in the detector, there are two energy regression GNNs, for either of these signatures. In the case of muons, their energy when passing through the detector is in the minimum ionisation range \cite{workman_22}, making it very hard to regress their energy, especially without a magnetic field, as is the case in this simulation. All pipelines have the same structure of repeated convolution and pooling layers, followed by a readout layer for the input of an MLP. Their exact definition slightly differ, most importantly by the choice of threshold for Treclus, either dataset specific or generalist (all particles at once), and in the granularity of the readout, chosen to be maximal in the regions of interest (whole detector, ECAL or HCAL) while not imposing an MLP size that would be too big. For all GNN pipelines, we apply 3 layers of convolution and pooling. At every CP layer, the number of node features doubles while the number of nodes is divided by two. For readout, the granularities depend on the task at hand. For the particle ID, we need a granular readout for the full detector, both longitudinally (for e$^-/\gamma$ and $\pi^+$ classification) and radially (to distinguish $\mu$, which do not spread radially in the detector). As such the ECAL is divided in 3 concentric region, while the HCAL  has 4 longitudinal regions, divided in respectively 3, 3, 2 and 2 concentric regions. For energy regression tasks, we divide the concerned part of the detector (ECAL or HCAL) with a granular readout (3 to 6 longitudinal slices, dependent on the length of the region, divided each in 3 concentric cylinders), while keeping the rest of the detector as a single readout region. Finally, the MLP is funnel-shaped comprised of 6 layers with leaky ReLU activation.
    		
    \section{Results}
    
\begin{figure}
    			\centering
    			\begin{minipage}{0.3\textwidth}
    				\centering
    				\includegraphics[width=\textwidth]{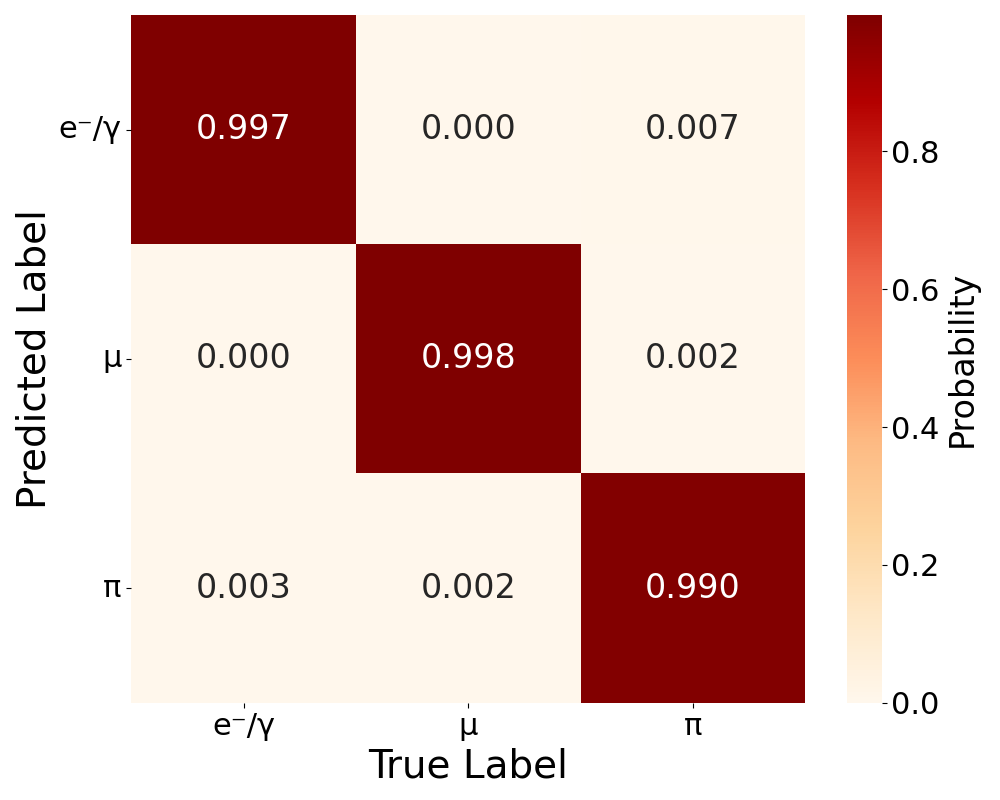}
    				\vspace{-0.2cm}
    				\captionof{table}{Normalised confusion matrix of the particle classifier pipeline.}\label{Table-confusion}
    			\end{minipage}
	    		\hfill
	    		\begin{minipage}{0.65\textwidth}
	    			\centering
	    			\begin{subfigure}{0.45\textwidth}
	    				\centering
	    				\includegraphics[width=\textwidth]{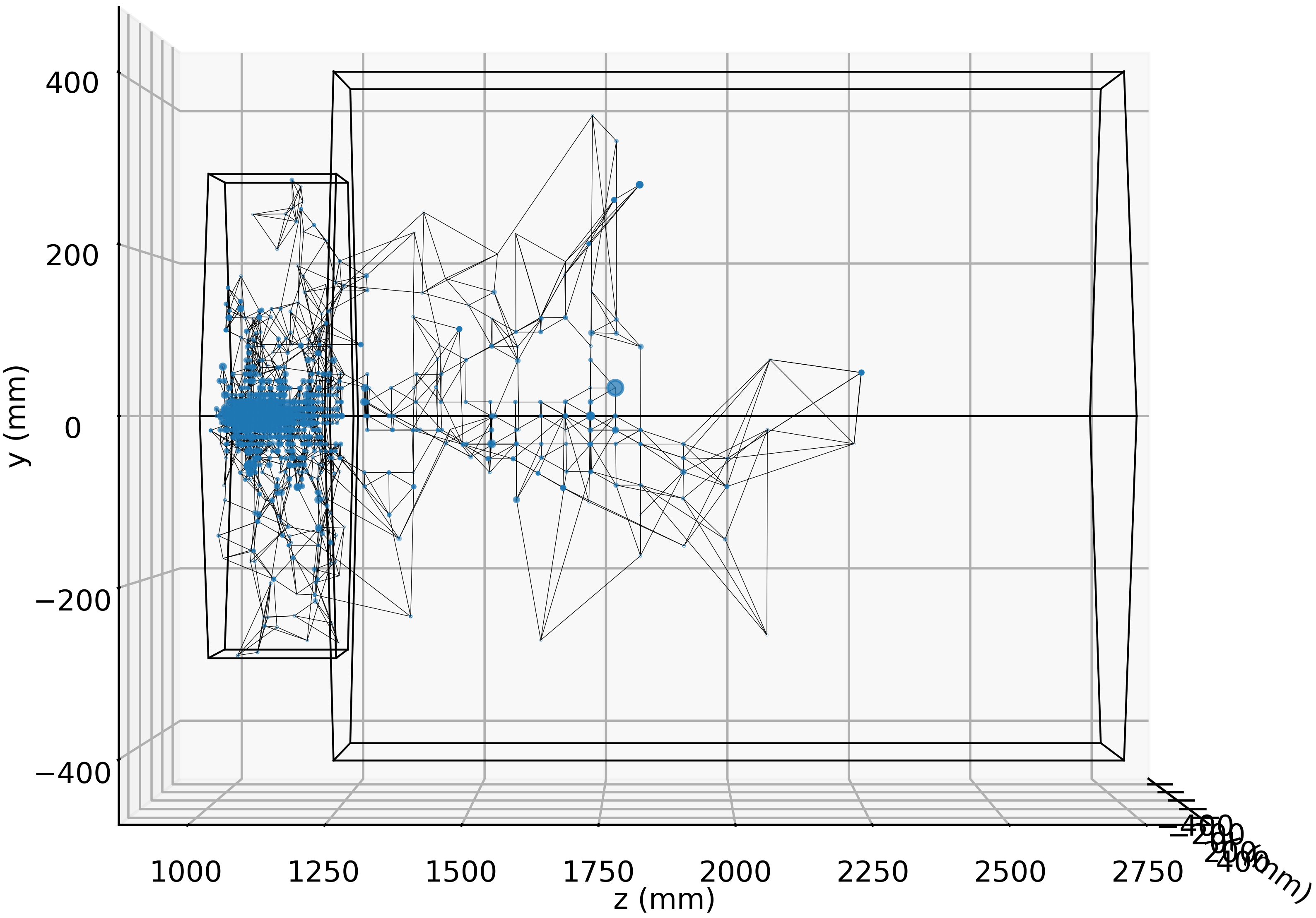}
	    				\caption{}\label{SubFig-hadronic_elm}
	    			\end{subfigure}
	    			\begin{subfigure}{0.45\textwidth}
	    				\centering
	    				\includegraphics[width=\textwidth]{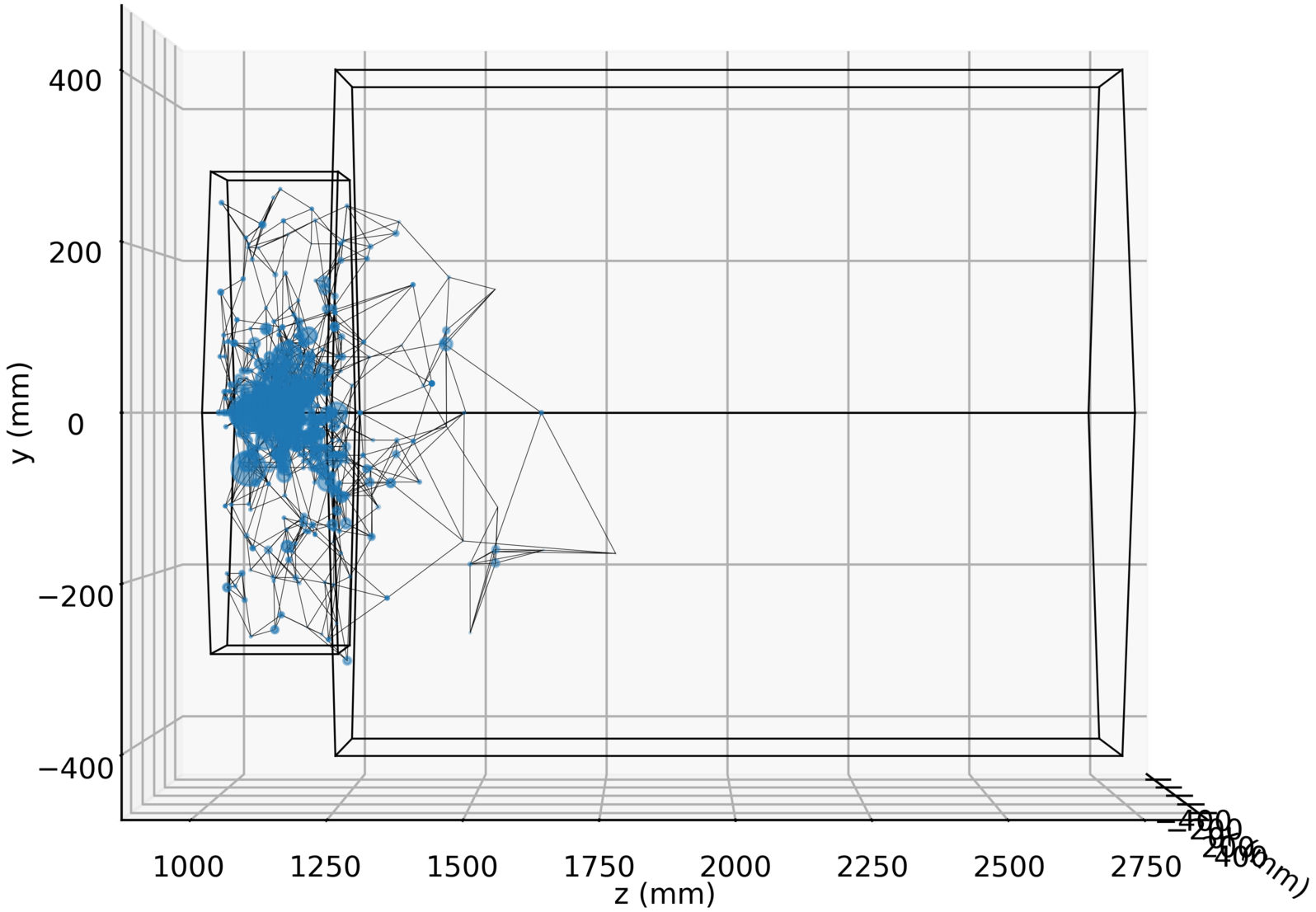}
	    				\caption{}\label{SubFig-em_pip}
	    			\end{subfigure}
	    			\vspace{-0.6cm}
	    			\captionof{figure}{Misclassified events with the typical signatures of the particles they are misidentified as. (\ref{SubFig-hadronic_elm}) An electron induced hadronic shower  (\ref{SubFig-em_pip}) An early showering pion ($\pi^+ \rightarrow \pi^0 \rightarrow \gamma \gamma$) \cite{wigmans_17}.}\label{Fig-misclassified}
	    		\end{minipage}
	    	\end{figure}

\begin{figure}
    			\centering
    			\begin{minipage}{0.64\textwidth}
    				\centering
	    			\begin{subfigure}{0.49\textwidth}
	    				\centering
	    				\includegraphics[width=\textwidth]{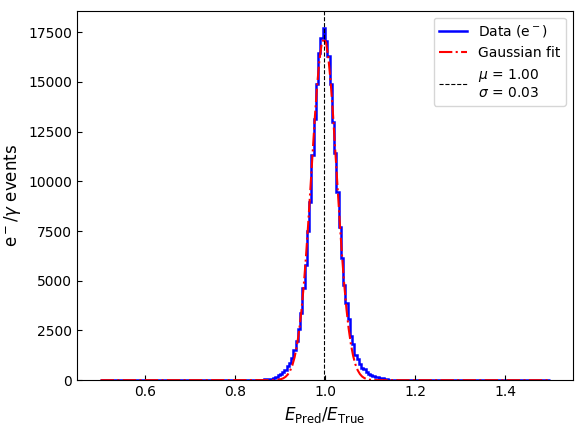}
	    			\end{subfigure}
	    			\begin{subfigure}{0.49\textwidth}
		    			\centering
		    			\includegraphics[width=\textwidth]{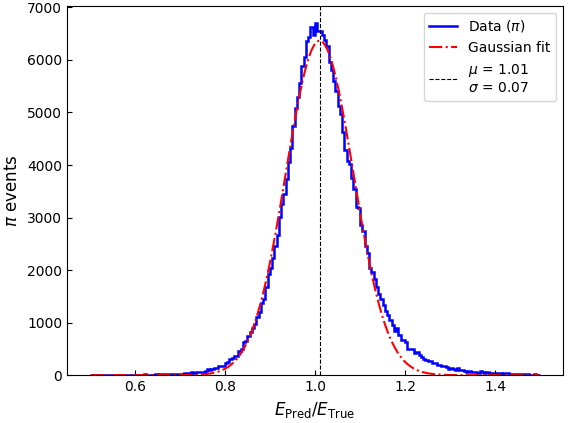}
	    			\end{subfigure}
	    			\vspace{-0.2cm}
	    			\captionof{figure}{Energy response $E_\text{pred} / E_\text{true}$.}\label{Fig-energy_ratios}
	    		\end{minipage}
	    		\hfill
	    		\begin{minipage}{0.35\textwidth}
    				\centering
    				\includegraphics[width=0.9\textwidth]{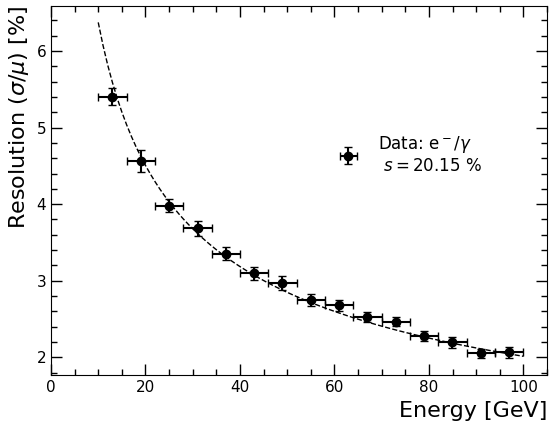}
    				\vspace{-0.4cm}
    				\captionof{figure}{e$^-$ energy resolution.}\label{Fig-resolution}
	    		\end{minipage}
    		\end{figure}

    	%\subsection{Training Parameters}
    		The models were trained on pipelines with 3 graph convolution layers and an MLP with 6 hidden layers, for a total of $\sim 10^4$ parameters. The training set for the classifier was composed of $5\times 10^5$ events. The training sets for either regression networks were trained from a set of $2\times 10^6$ events, labelled as either e$^-/\gamma$ or $\pi^+$ by the previously trained classifier ($\sim 10^6$ graphs per regressor).
   % 	\subsection{Particle Identification}
    		We first look at the classifying performance of the GNN pipeline presented above. As can be guessed from Figure \ref{Fig-event_displays}, this is mostly a density problem and the choice of readout regions is the key element of this task. From the confusion matrix (table \ref{Table-confusion}) we can see a good classification performance.
    		If we look at the raw detector output of misclassified elements (figure \ref{Fig-misclassified}), we observe that most errors are due to outlier events, such as electromagnetic showers leaking into the HCAL, electron induced hadrons, later detected in the HCAL or early showering pions.
    %	\subsection{Energy Regression}
    		The training of the energy regression pipelines was done with events classified by the Particle ID pipeline as either e$^-/\gamma$ or $\pi^+$. We define the response as the ratio $E_{\text{pred}} / E_{\text{true}}$ and use it as a metric for performance. We observe a Gaussian-like distribution centred around 1 (see figure \ref{Fig-energy_ratios}).  The difference in ECAL and HCAL sampling fractions mean electromagnetic showers have a smaller standard deviation than for pions.
    		%We also look at the energy resolution
    		%\begin{equation}
    		%	\left( \frac{\sigma \left( \frac{E_{\text{pred}}}{E_{\text{true}}} \right)}{\left\langle \frac{E_{\text{pred}}}{E_{\text{true}}} \right\rangle} \right)^2 = \left( \frac{S}{\sqrt{E_{\text{true}}}} \right)^2 + \left( \frac{N}{E_{\text{true}}} \right)^2 +  C^2
    		%\end{equation}
    		%with $S$ the stochastic term, describing the physical fluctuations in shower development, $N$ the noise term, describing the electronic noise from the detector readout electronics and $C$ the constant term, describing systematic noise \cite{fabjan_03}.
    		We also look at the energy resolution $\left.\sigma \left( \frac{E_{\text{pred}}}{E_{\text{true}}} \right) \right/ \mu \left( \frac{E_{\text{pred}}}{E_{\text{true}}} \right)$ as defined in \cite{fabjan_03}. In our simulation, no readout or systematic noise have been emulated and we therefore only consider $\frac{\sigma}{\mu} \propto \frac{1}{\sqrt{E}}$, as seen in figure \ref{Fig-resolution}. For the electromagnetic calorimeter, we find a stochastic term $S_{\text{GNN}} = 20.15\%$, close to the theoretical value $S_{\text{exp}}$ of 20\%.

    \section{Conclusion}
    
    	%\subsection{Outlook}
    		Our framework, by its generic and detector agnostic nature, can be generalised to other problems in high granularity particle detectors with a non-regular architecture, such as diffuse supernova neutrino background searches in Super and Hyper-Kamiokande \cite{beacom_04} or overlapping object segmentation in granular detectors (HGCAL or Hyper-Kamiokande like).
    	%\subsection{Summary}
	    	We presented a message passing graph convolution neural network architecture for particle identification and energy regression in high granularity calorimeters. We introduced the use of Proximity Tables and the PT-KNN algorithm for optimised graph construction. In particular, we discussed the improved average time complexity compared to the classical KNN algorithm. We also discussed the possibility to reduce the size of proximity tables for PT-KNN implementation in resource constrained environments such as on FPGAs. We also presented our implementation of message passing convolution and pooling layers. Treclus, our threshold based clustering algorithm for geometric pooling of graph has proven to be very adapted to coarsen graphs while preserving the original global structure of the graph. We discussed the need for a graph readout pooling to feed the result of the convolution layers to an MLP. We implemented a segmentation  of the detector based on the geometry of the physical processes at hand to flatten the graph structure in a fixed size tensor while preserving the symmetries of the problem. Finally, we discussed the state of the art level classification and energy regression capacities of our GNN implementation. Looking at the energy resolution of our algorithm, we recover the expected values, proving the validity of our approach.
	    	
	\section{Acknowledgements}
		We are grateful for the Agence Nationale de la Recherche, funding the OGCID project under funding ANR-21-CE31-0030. 
		
	\printbibliography
	
\end{document}